# Quantum Phase Slips in Superconducting Nanowires

C. N. Lau, N. Markovic, M. Bockrath, A. Bezryadin[†] and M. Tinkham

Department of Physics, Harvard University, Cambridge, MA 02138

[†]Present address: Department of Physics, University of Illinois at Urbana-Champaign, Urbana, IL 60801.

**We have measured the resistance vs. temperature of more than 20 superconducting nanowires with nominal widths ranging from 10 to 22 nm and lengths from 100 nm to 1 µm. With decreasing cross-sectional areas, the wires display increasingly broad resistive transitions. The data are in very good agreement with a model that includes both thermally activated phase slips close to $T_C$ and quantum phase slips (QPS) at low temperatures, but disagree with an earlier model based on a critical value of $R_N/R_q$. Our measurements provide strong evidence for QPS in thin superconducting wires.**

According to the Mermin-Wagner theorem[1], superconducting long-range order is impossible in a strictly one-dimensional system. Here we ask how superconductivity is extinguished as a superconducting wire is made narrower and narrower. Although thermally-activated phase slips dominate near the superconducting transition temperature ($T_c$), our measurements confirm the dominant role of quantum phase slips below $T \sim T_c/2$.

Phase slips give rise to resistance in a thin superconducting wire below $T_C$. During a phase slip, the superconducting order parameter fluctuates to zero at some point along the wire, allowing the relative phase across the point to slip by $2\pi$, resulting in a voltage pulse; the sum of these pulses gives the resistive voltage. In a theory developed by Langer, Ambegaokar, McCumber and Halperin (LAMH)[2], such phase slips occur via thermal activation as the system passes over a free-energy barrier proportional to the cross-sectional area of a wire. Experiments on 0.5 μm diameter tin whiskers confirmed the theory[3].

Subsequently, Giordano[4] observed in thin In and PbIn wires a cross-over from the LAMH behavior near $T_C$ to a more weakly temperature dependent resistance tail at lower temperatures. He attributed this tail to phase slips occurring via macroscopic quantum tunneling, or quantum phase slip (QPS), through the same free-energy barrier. However, interpretation of these results has been complicated by the possibility of granularity in these metals that could give rise to a similar temperature dependence. Additionally, Sharifi *et al*[5] found in homogenous Pb wires a systematic broadening of the transition with decreasing cross-sectional areas of the wires that could not be explained by the LAMH theory, but no cross-over to a more weakly temperature dependent regime was observed. Thus it is controversial whether such quantum phase slips have been observed in experiments. Theoretically, it is also a subject of debate whether resistance arising from QPS is actually observable[6-9]. Certain theories imply that QPS should be important when the wire diameter is about 10nm[7], but such thin wires are extremely difficult to fabricate by conventional electron beam lithography.

To address this question, we have developed a new fabrication technique[10]. In this Letter we report measurements of a large number of amorphous MoGe wires with various widths and lengths. A systematic broadening of the superconducting transition with decreasing cross-sectional areas is observed, which can be described quantitatively by a combination of thermally activated phase slips close to $T_C$ and QPS at low temperatures. Using a simple model with only two free parameters of order unity for the entire family of curves, surprisingly good fits of the data over a wide range of samples are obtained, thus providing convincing experimental evidence for quantum phase slips.

The nanowires are fabricated by sputtering 4 – 5 nm of $Mo_{0.79}Ge_{0.21}$, followed by 1 – 2 nm of germanium (as a protective layer against oxidation), onto carbon nanotubes or ropes which are suspended across slits on $SiN/SiO_x/Si$ substrates[10]. The slits have widths 100, 150, 250, 350 and 550 nm. The wires we measured have apparent widths ranging from 10 to 22 nm, and lengths 100 nm to 1 μm, as determined from scanning electron microscope (SEM) images. These nominal widths are overestimates because the Ge protective layer is included in the image, and because of the resolution limit of the SEM.

One concern in studying ultrathin wires is possible granularity in samples. However, our MoGe wires are believed to be homogeneous and non-granular because: (1) it is known that MoGe is amorphous and can be electrically continuous down to 1 nm in thickness[11], (2) the wires' measured $R_N$ do not differ significantly from that calculated from their geometry and bulk resistivity values, and (3) TEM images of wires prepared under similar conditions did not reveal any granularity[10]. Fluctuations in the sample width are about 1 nm in size as seen in TEM images. We also believe that the underlying nanotubes do not contribute to the *dc* conductance of the nanowires[12-14], because the nanowires' measured normal resistances agree with that calculated from their geometry and bulk resistivity values with no correction from the nanotubes.

The samples' resistances are obtained from measured slopes of the current-voltage characteristics. The biasing current $I_B$ had a frequency of 0.1 Hz and amplitude of 3nA. Doubling or halving the amplitude did not change the slope, showing that $I_B$ was indeed within the linear part of the I-V curve. Over 20 samples were measured, and Fig. 1 shows the resistance vs. temperature curves for representative samples (those not shown here have similar behaviors and are omitted to reduce clutter). For each curve, the first sharp drop is due to the superconducting transition of the co-evaporated thin film electrodes, which were unavoidably included in the measurements of the nanowires and underwent a sharp transition at about 5 – 5.5K. Since transitions of the wires occur at lower temperatures and are considerably broader, the measured resistance of a sample below the film $T_C$ can be attributed solely to the nanowire. In particular, the normal state resistance of the wire ($R_N$) is taken to be the measured resistance just below the film $T_c$. Note that any proximity effect on a wire from the superconducting banks is not significant, since Cooper pairs can only diffuse a distance $\xi_N$ into a dirty normal metal, where $\xi_N = \sqrt{hD/2\pi kT} < 8$ nm for MoGe, much shorter than our wires. (D=0.5 cm$^2$/s is the diffusion constant[11].)

Our previous measurements of wires roughly 150 nm long suggested that the wires were superconducting (resistances decreased rapidly) if their *total* normal state resistance $R_N < R_q$, and insulating/metallic (resistances barely changed with temperature) if $R_N > R_q$, where $R_q$=6.5 k$\Omega$ is the quantum resistance for pairs[10]. In contrast to this apparent simple dichotomy in the previous results, the R-T curves in Fig. 1 display a broad spectrum of behaviors, including some superconducting samples with resistance as high as 40 k$\Omega$ (>>$R_q$). These data on a more comprehensive set of samples lead to a different conclusion from the previous results, since it indicates that the relevant parameter controlling the superconducting transition is not the ratio of $R_q/R_N$, but appears to be resistance per unit length, or equivalently, the cross-sectional area of a wire. This is illustrated by the solid lines in Fig. 2, which plots R/L vs. t≡T/T$_{c,film}$, the temperature normalized to film $T_c$. Here resistances of wider wires ($R_N/L < 20$ $\Omega$/nm) drop relatively sharply below $T_{c, film}$, whereas the transition widths broaden with increasing values of $R_N/L$, and resistances of the narrowest wires ($R_N/L >80\Omega$/nm) barely change with temperature down to 1.5

K.

To understand this systematic broadening of the transitions of the wires with decreasing cross sectional area A, we first consider the LAMH theory, which explains resistive transitions in terms of proliferation of thermally activated phase slips over a free-energy barrier $\Delta F$ proportional to A. This leads to a resistance below $T_C$

$$R_{LAMH} = \frac{\pi h^2 \Omega}{2e^2 kT} e^{-\Delta F/kT}, \qquad (1)$$

where $\Omega = \frac{L}{\xi}\left(\frac{\Delta F}{kT}\right)^{1/2}\frac{1}{\tau_{GL}}$ is the attempt frequency, and $\Delta F = \frac{8\sqrt{2}}{3}\frac{H_c^2}{8\pi}A\xi$ is the energy barrier. In these expressions, L is the length of the wire, $H_c$ and $\xi$ are the thermodynamic critical field and the coherence length of the material, T is the temperature, k is the Boltzman constant, and $\tau_{GL} = \frac{\pi h}{8k(T_C - T)}$ is the characteristic relaxation time in the time-dependent GL theory. Eq. (1) predicts negative curvature of logR(T) for all values of T and unmeasurably small resistances for $t = T/T_C < 0.3$ even for the narrowest wires we measured; neither of these predictions agree with the majority of the data.

This discrepancy between predictions of LAMH theory and our data leads us to consider the possibility of MQT of phase slips. A heuristic argument due to Giordano[4] suggests that the resistance from MQT follows a form similar to (1), except that the appropriate energy scale is $h/\tau_{GL}$ instead of kT. Therefore, we expect

$$R_{MQT} = B\frac{\pi h^2 \Omega_{MQT}}{2e^2 (h/\tau_{GL})} e^{-a\Delta F \tau_{GL}/h}, \qquad (2)$$

where $\Omega_{MQT} = \frac{L}{\xi}\left(\frac{\Delta F}{h/\tau_{GL}}\right)^{1/2}\frac{1}{\tau_{GL}}$, and *a* and *B* are possible numerical factors of order unity. MQT causes phase slips even as T 0 and results in experimentally measurable resistance at all temperatures for sufficiently narrow wires. Therefore the total resistance in the superconducting channel will be $R_{LAMH} + R_{MQT}$. However, unless this is small compared to $R_N$, current carried by the parallel normal channel will significantly reduce the observed resistance. To take account of this in a simple way, we take the total resistance predicted by our model to be

$$R = \left[ R_N^{-1} + (R_{LAMH} + R_{MQT})^{-1} \right]^{-1}. \tag{3}$$

To compare Eq. (3) directly with the data, we note that the dominant exponential terms are determined by the cross sectional area A, which can be conveniently described by a dimensionless parameter c relating the energy barrier for phase slips to thermal energies near $T_C$,

$$c \equiv \frac{\Delta F(T=0)}{kT_c} = \frac{8\sqrt{2}}{3} \frac{H_C(0)^2}{8\pi} \frac{\xi(0)}{kT_c} A. \tag{4}$$

Using standard BCS and Ginzburg-Laudau relations for dirty superconductors[15], this expression can be re-written in terms of parameters that are more experimentally accessible,

$$c \approx 0.83 \frac{R_q}{\xi(0)} \frac{L}{R_N} = 0.83 \frac{R_q}{R_{\xi(0)}}, \tag{4a}$$

where we have introduced the notation $R_{\xi(0)}$ to refer to the normal resistance of the wire in a length $\xi(0)$. For the samples reported here, taking $\xi(0)$=5nm[11], (4a) yields values of c ranging from 8 to 39.

Using these values of c, we calculate the resistance of the wires arising from both thermal and quantum phase slips, as given in Eq. (3), with two adjustable parameters, *a* and *B*[16]. (The calculated curves are also multiplied by an overall factor 1.2 so that they can be compared more easily with the data.) As shown by the dotted lines in Fig. 2, these simulations reproduce the data quite well when we take *a*=1.3 and *B*=7.2. The agreement is rather remarkable since there are only two free parameters for the *entire family* of curves.

The above model is based on a heuristic formulation. However, in a recently developed microscopic theory by Golubev and Zaikin[8], the MQT term follows an exponential term identical to that in (2) (apart from factors of order unity), but the prefactor has an additional factor of $a\sqrt{c}/0.83$, which is about 7 on average for our samples. This different prefactor given by the microscopic theory provides a good explanation for the somewhat large value of *B* obtained from our fits. Moreover, by introducing small random fluctuations in the width (i.e. the values of c) along a single wire, we are able to reproduce the occasional crossing of the data curves as seen in samples 6 and 7.

We note that Eq. (3) cannot reproduce the data of sample 5 (and one other sample not shown here) adequately for any choice of *a* and *B*. This may be attributed to a number of mechanisms, such as depressed $T_C$[17], unusually thin films or inadvertent contamination. Nevertheless, since only 2 of the 20 samples show such anomalous behaviors, and since *a*, the factor of order unity in the dominant exponential term, is found through simulations to be within 30% of unity, the agreement between the majority of the data and the model (3) is still quite remarkable.

This simple model considers only individual non-interacting quantum phase slips. This is supported by the theoretical work of Golubev and Zaikin[8], which argues that interactions should not be important except in considerably longer wires than those studied here.

To address the question of whether there is a well-defined cutoff diameter, below which superconductivity is excluded even at T = 0, we plot the normalized sample resistances at our lowest temperatures (~1.5K) as a function of $L/R_N$ in Fig. 3. (The parameter $L/R_N$, rather than cross-sectional area A is used because L and $R_N$ are known with much greater accuracy than the widths and profiles of the wires, which would be needed to determine A geometrically.) The linear plot in Fig. 3a suggests that there is a transition from metallic to superconducting states at $L/R_N \sim 0.014$ nm/Ω, corresponding to a sample width of about 10 nm. This is numerically consistent with theoretical predictions of a critical width ~10 nm[7]. However, the significance of this agreement is unclear in view of Fig. 3b, which plots the same data points on a semi-log scale and demonstrates that there is no feature at any particular value of $L/R_N$. In fact, Fig. 3b can be readily understood in terms of MQT of phase slips at low temperatures. The plot shows that resistances at $T/T_C \sim 0.3$ decrease exponentially with $L/R_N$. This is what we would expect from equation (2) since $\Delta F \propto A \propto L/R_N$, and the contribution (1) from thermally-activated phase slips is negligible at such low temperature. Quantitatively, if we neglect the weak effect of the prefactor and consider only the exponential dependence in (2), the slope of the semi-log plot is calculated to be

$$\frac{\partial \ln(R_{MQT}/R_N)}{\partial (L/R_N)} = a \cdot 0.83 \frac{\pi}{8} \frac{R_q}{\xi(0)} \approx 0.54 k\Omega/nm. \qquad (5)$$

where we have used $a=1.3$ obtained from the simulations. Fitting the data points in Fig. 3b with an exponential function, we obtain a slope of 0.39 kΩ/nm, in reasonable agreement with the expected value (5). Therefore our simple model of quantum phase slips works remarkably well in explaining both the exponential form and numerical coefficient of the dependence of the data on sample cross-section.

Finally, we would like to comment on the role of dissipation. Dissipation from a shunt resistor below $R_q$ is predicted to suppress MQT of phase slips in Josephson junctions[18, 19]. In our nanowires, presumably the primary source of dissipation is the loss from high-frequency electromagnetic fields generated in the nanowire by the phase-slips. At the low temperature limit, from Eqs. (2) and (4a), we see that

$$R_{MQT} \propto \exp-(bR_q/R_{\xi(T)}), \qquad (6)$$

where $b \approx 0.43$. This result is reminiscent of the Schmid result for Josephson junctions[18], if one takes the damping resistance to be the normal resistance of a few coherence lengths of the nanowire surrounding the phase slips. This conclusion is consistent with the work of Golubev and Zaikin[8], who point out that dissipation, while physically important, does not appear in the final formulae for phase slip rates apart from numerical factors of order unity. An additional source of dissipation could stem from the capacitive coupling at high frequencies between the nanowire and the underlying nanotubes. This would depend on the fraction of the nanotubes in the ropes that were metallic, and hence dissipative. However, we would like to stress that Eq. (3) fits the data remarkably well, without invoking any role for external dissipation.

We thank B.I. Halperin and E. Demler for helpful discussions, S. Shepard for assistance with fabrication, and R. Smalley for providing the nanotubes. This work is supported by NSF grants DMR-00-72618, DMR-98-0936, PHY-98-71810 and ONR grant N00014-96-0108.

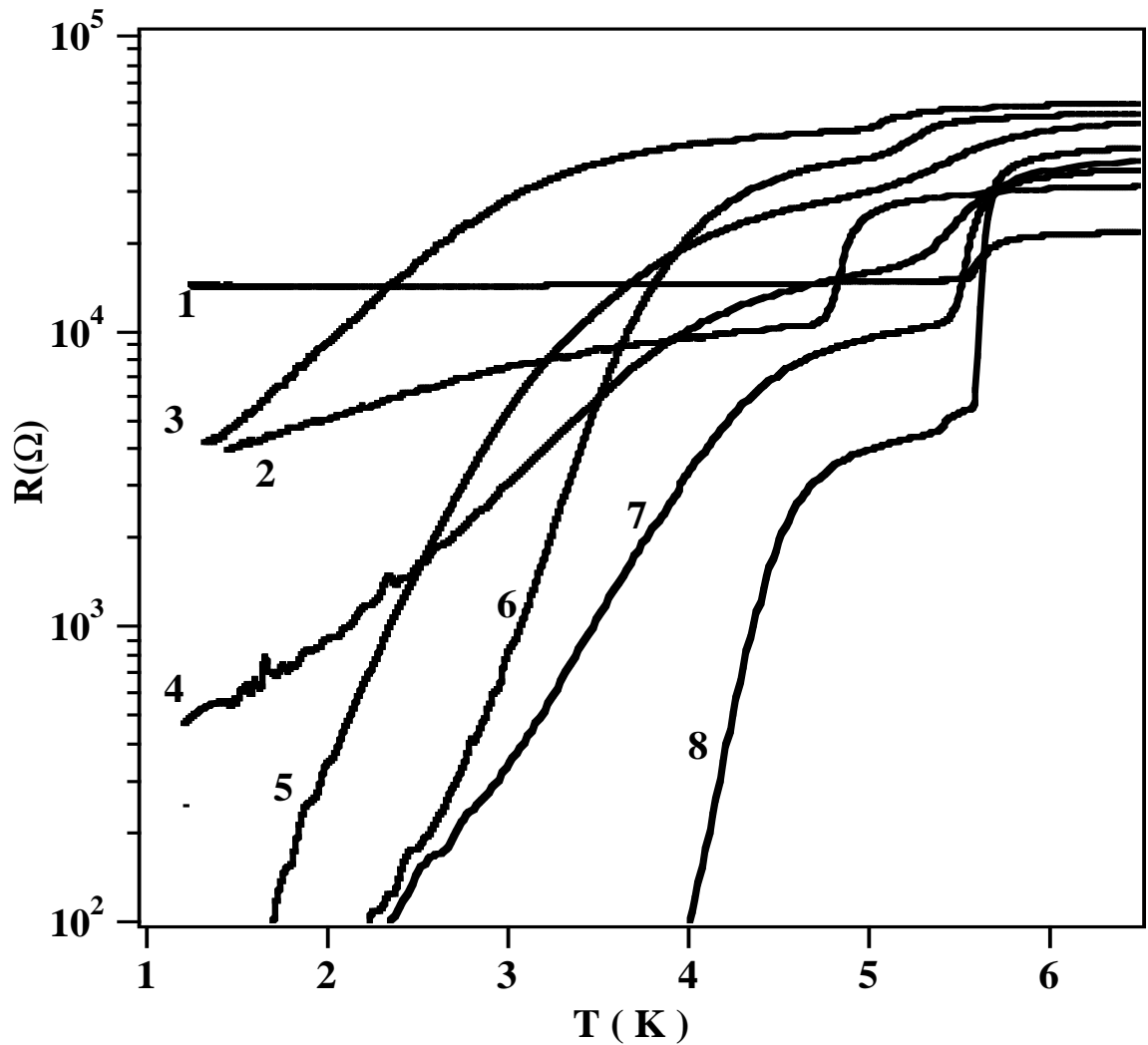

**Fig. 1.** Resistances as a function of temperature for 8 different samples. The samples' normal state resistances and lengths are: 1: 22.6 kΩ, 185 nm; 2:10.7 kΩ, 135 nm; 3: 47 kΩ, 745 nm; 4:17.3 kΩ, 310 nm; 5: 32kΩ, 730 nm; 6: 40 kΩ, 1050 nm; 7:10 kΩ, 310 nm; 8:4.5 kΩ, 165 nm.

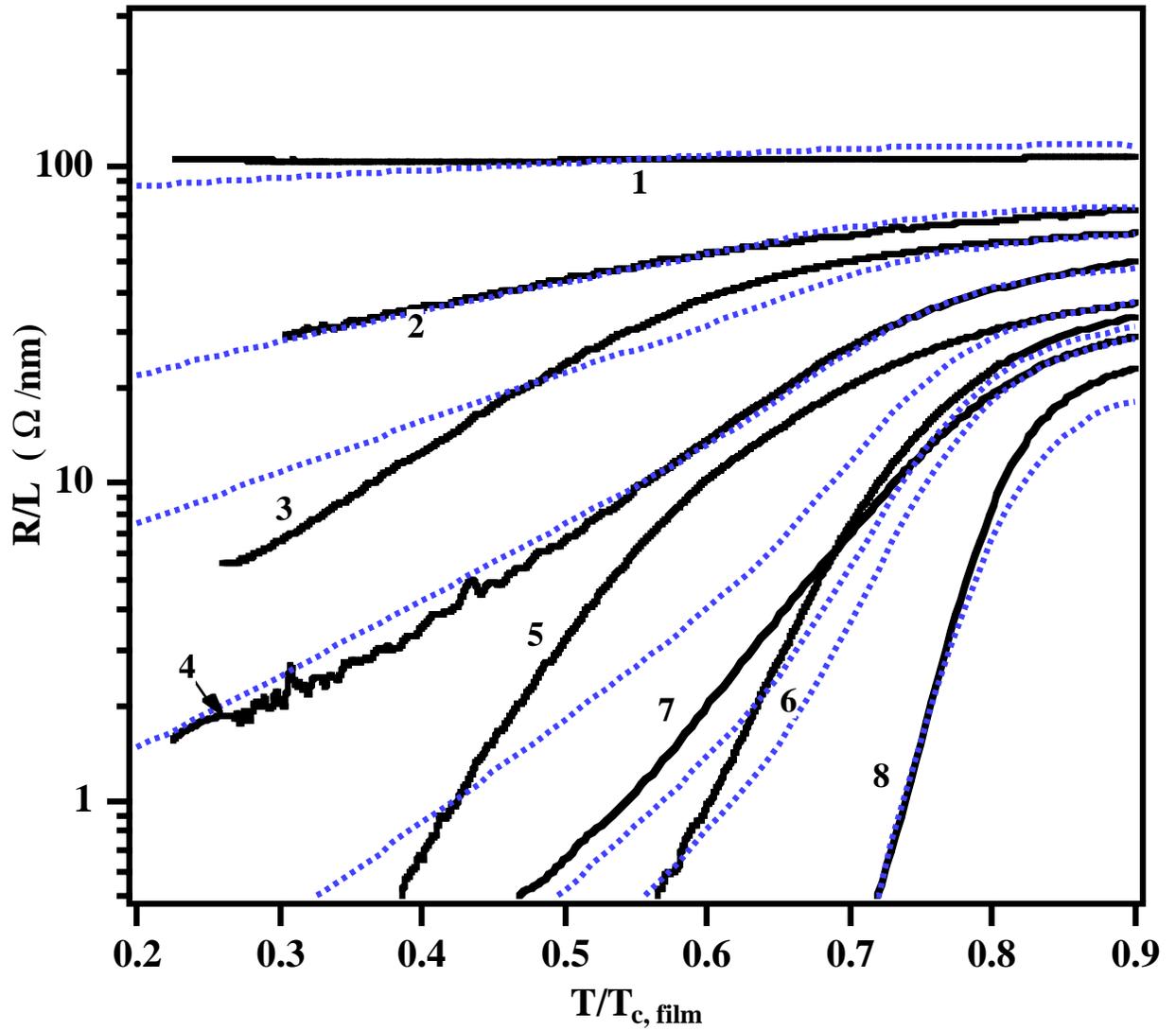

**Fig. 2**. The solid lines are the data showing the measured resistance per unit length vs. normalized temperatures. The dotted lines are curves calculated using Eq. (3) and sample parameters. The two free parameters used are $a$=1.3 and $B$=7.2 for the whole family of the curves.

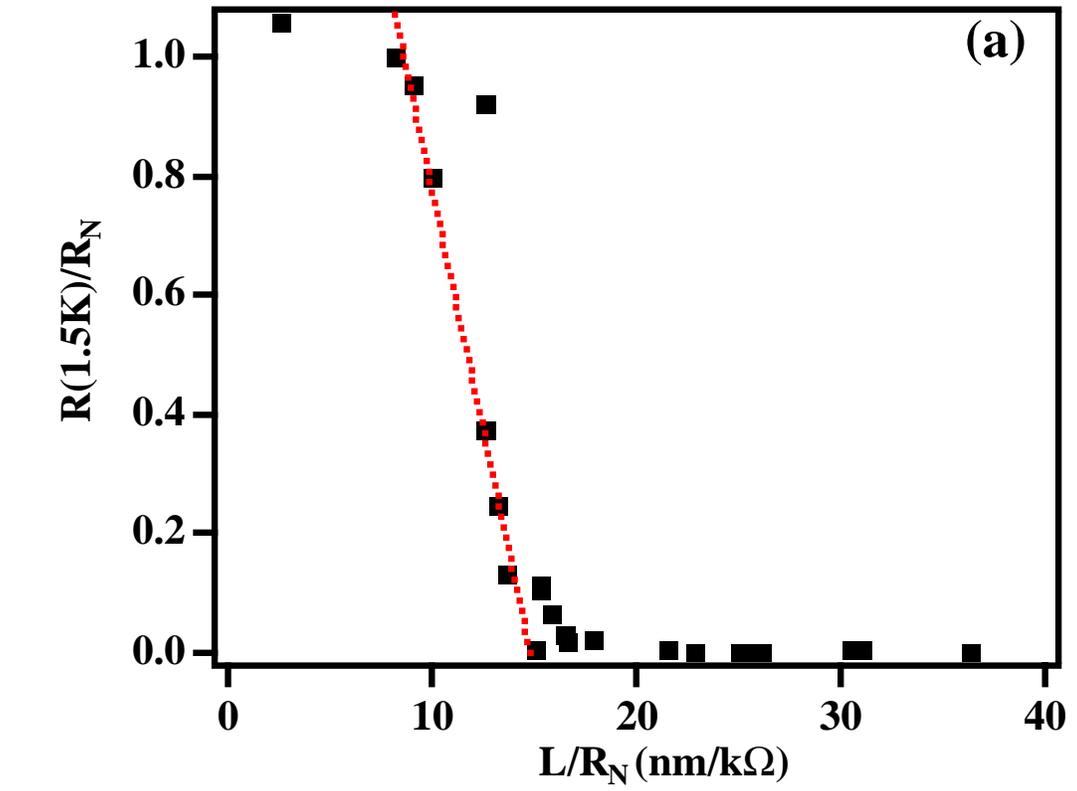

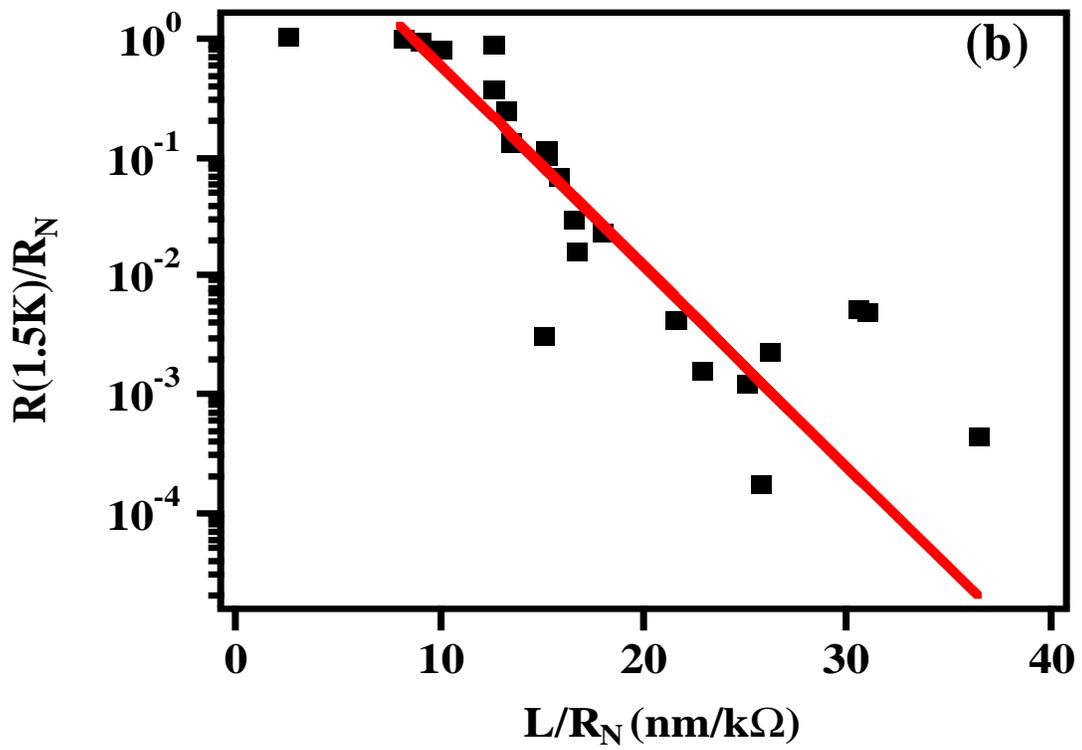

**Fig. 3.** Resistance at 1.5 K normalized to normal state resistance as a function of $L/R_N$. (a) Linear plot. The dotted line is a guide to the eyes. (b) Semi-log plot with an exponential fit. Slope of the fitted line is 0.39 kΩ/nm, compared with 0.54 kΩ in (5).